# Enhanced thermoelectric performance of twisted bilayer graphene nanoribbons junction

Shuo Deng[1,2], Xiang Cai[2], Yan Zhang[3*], and Lijie Li[2*]
[1]School of Logistic Engineering, Wuhan University of Technology
[2]College of Engineering, Swansea University, Swansea, SA1 8EN, UK
[3]School of Physics, University of Electronic Science and Technology of China
*Emails: zhangyan@uest.edu.cn, L.Li@swansea.ac.uk

**Abstract**

We investigate the electron transport and thermoelectric property of twisted bilayer graphene nanoribbon junction (TBGNRJ) in 0º, 21.8º, 38.2º and 60º rotation angles by first principles calculation with Landauer-Buttiker and Boltzmann theories. It is found that TBGNRJs exhibit negative differential resistance (NDR) in 21.8º and 38.2º rotation angles under $\pm 0.2$ V bias voltage. More importantly, three peak *ZT* values of 2.0, 2.7 and 6.1 can be achieved in the 21.8º rotation angle at 300K. The outstanding *ZT* values of TBGNRJs are interpreted as the combination of the reduced thermal conductivity and enhanced electrical conductivity at optimized angles.

## 1 Introduction

Graphene is the first true 2-dimensional (2D) material, which consists of carbon atoms forming regular hexagonal lattice[1]. In the prior research, many outstanding properties of graphene were discovered, such as high electrical conductivity and carrier mobility[2-6], high thermal conductivity[7], and superior mechanical properties[8, 9]. Hence graphene can have many applications in electronic and mechanical devices. In particular, it has been applied in thermoelectric devices[10, 11]. The energy conversion efficiency of thermoelectric materials can be described by the figure of merit $ZT = S^2 G_e T/(\kappa_e + \kappa_{ph})$, where the $G_e$, $S$ and $T$ are the electrical conductance, Seebeck coefficient and temperature, respectively. $\kappa_e$ and $\kappa_{ph}$ are the heat transport coefficient of electrons and phonons. A good thermoelectric material should have a high electrical conductivity, Seebeck coefficient and low thermal conductivity. On one hand, the Dirac-cone band structure of graphene makes it to display a high electrical conductivity[5, 6]. On the other hand, the thermoelectric performance of graphene is poor because of its high thermal conductivity, and the closed bandgap which leads to a small Seebeck coefficient[12]. In fact, these properties cannot change independently because they correlate with each other[13]:

$$S = \frac{8\pi^2 k_\beta^2}{3eh^2} m^* T \left(\frac{e\mu\pi}{3G_e}\right)^{2/3} \qquad (1)$$

where, $k_\beta$, $e$, $h$, $m^*$, $n$ and $\mu$ are the Boltzmann constant, electron charge, Planck's constant,

effective mass, carrier concentration and mobility, respectively. From the equation (1), the Seebeck coefficient is inversely proportional to the electrical conductance. As a result, the key to improve the *ZT* of graphene devices is find a trade-off between the Seebeck coefficient, electrical conductance and heat transport coefficient. In prior studies, the specially designed nanostructured graphene can have increased Seebeck coefficient and suppressed heat transport coefficient without greatly reducing electrical conductance due to the quantum confinement effect[10, 14, 15]. In these graphene devices, graphene nanoribbons (GNRs) has demonstrated better thermoelectric performances as the Seebeck coefficient and *ZT* value can be increased by the finite size effect[16-19]. One of earlier attempts on the thermoelectric performance of GNRs was reported by Ouyang *et al*[19]. They found that a higher Seebeck coefficient in GNRs compared with the pristine graphene is attributed to the edge geometry of GNRs, which plays an important role in enhancing the thermoelectric performance. In 2012, Jin *et al* reported that the armchair GNR exhibits a higher *ZT* value than zigzag GNRs, and *ZT* value increases with the decrease of the GNRs width[16]. A chevron type edge GNR was proposed demonstrating a *ZT* value of 3.25 at 800 K[20]. Huang *et al* discovered that the thermal conductance in the bilayer GNRs is hundreds times weaker than the pristine graphene because the weak van der Waals (vdW) interaction between two layers[21]. In recent several years, in order to improve thermoelectric performance of GNRs, different schemes based on more sophisticated GNR nanostructures have been designed, such as using boundary effect[17, 22, 23], interface effect[24], doping[18, 25], and structure defect or wrinkles[26-29].

Bilayer graphene structures with optimized twisting angles aiming to achieve much improved performances such as high temperature superconductivity and strong interlayer coupling have been theoretically and experimentally investigated recently[30-33]. Because the bilayer graphene structure efficiently limits the thermal conductivity in the normal direction to the 2D plane attributed to the vdW force[21], this will lead to a new approach to further increase the thermoelectric performance.

In this work, the electron transport and thermoelectric performance of twisted bilayer graphene nanoribbon junction (TBGNRJs) in 0°, 21.8°, 38.2° and 60° rotation angles are systematically researched using the first principles method. It is discovered that that TBGNRJs exhibit negative differential resistance (NDR) in 21.8° and 38.2° rotation angles under $\pm 0.2$ *V* bias voltage. Moreover, *ZT* values of 2.0, 2.7 and 6.1 can be achieved at different chemical potentials for the 21.8° rotation angle at 300K.

## 2 Computational procedure

The theoretical simulation of the twisted bilayer graphene structures starts with the Generalized Lattice Match (GLM) method, which is for investigating the relationship between the mismatch strain and the number of atoms[34]. The optimization of twisting angles is conducted by using the balanced mismatch strain and number of atoms, it is targeted to have a lower mismatch strain and at the same time a smaller number of atoms. As shown in the Figure 1(a) and (b), the vectors $\mathbf{v}_1$ and $\mathbf{v}_2$ defined the surface cell of the bottom graphene layer while the vectors $\mathbf{u}_1$ and $\mathbf{u}_2$ as the surface cell of the top graphene layer. The relationship between [$\mathbf{v}_1$, $\mathbf{v}_2$] and [$\mathbf{u}_1$, $\mathbf{u}_2$] is expressed as $\mathbf{A}[\mathbf{u}_1, \mathbf{u}_2] = [\mathbf{v}_1, \mathbf{v}_2]$, where $\mathbf{A}$ is the affine transformation matrix. The rotation matrix $\mathbf{U}$ has a form as[34].

$$\mathbf{U} = \begin{bmatrix} \cos(\phi) & -\sin(\phi) \\ \sin(\phi) & \cos(\phi) \end{bmatrix} \quad (2)$$

$$\phi = |\phi_a - \phi_b|/2 \quad (3)$$

$$\mathbf{P} = \mathbf{U}^T \mathbf{A} \quad (4)$$

where, $\phi$ is the twisting angle of bilayers graphene, $\phi_a$ is the angle between the vectors $\mathbf{u}_1$ and $\mathbf{u}_2$, $\phi_b$ is the angle between the vectors $\mathbf{v}_1$ and $\mathbf{v}_2$, matrix $\mathbf{P}$ defines the 2D strain tensor for deforming one cell into the other. We only calculate the twisting angles from 0° to 60° because the crystal structure of graphene is a regular hexagon. In order to sustain the structural stability and avoid having a large size cell, we analyzed the twisting angles of 0°, 21.8°, 38.2° and 60°, at which the strain of lattice mismatch is calculated to be 0%. Previous studies have shown that AA (0° rotation angle) and AB (60° rotation angle) stacking of bilayer graphene are the most common stacking style in the research[35]. The other twisting angles (21.8° and 38.2°) were chosen, because they had a small size of cell comparing with other rotation angles.

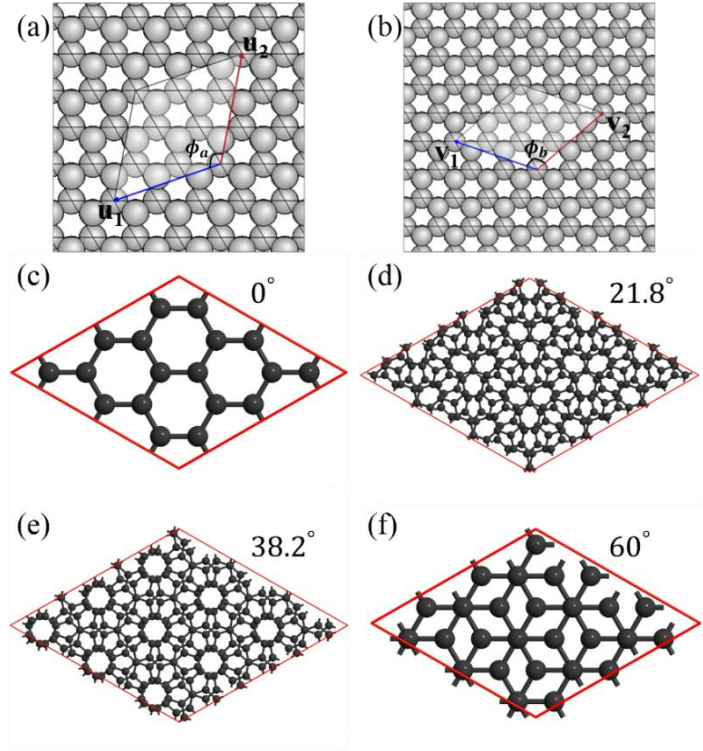

Figure 1. The surface cell of top (a) and bottom of graphene (b). (c)-(f) show the top view of the atomic arrangement of the twisted bilayer graphene with 0°, 21.8°, 38.2° and 60° rotation angles, respectively.

All the modelling and calculations have been implemented by the Quantum Atomistix ToolKit (ATK2018) simulation tools[36]. Calculations of the electronic properties have been performed within the framework of density functional theory (DFT). In the geometry optimization, we use the semi-empirical corrections by the Grimme DFT-D2 model to calculate the distance between two layers, which accounts the long-range vdW interaction[37]. The structure is fully relaxed until the force on each atom becomes smaller than 0.01 eV/Å. The generalized gradient approximation (GGA) with the parametrization of Perdew-Burke-Ernzerhof (PBE), mesh cut-off energy of 400 eV and 12×12×1 k-points grid were used. To avoid the interaction of the periodic boundary conditions, a large vacuum spacing of at least 25 Å is added along the normal direction to the electrons transport plane. The calculated rotation angle ($\theta$), lattice constant ($a$), mean absolute strain ($\varepsilon$) and surface distance between two layers ($d$) are listed in Table 1. For AA (0 ° rotation angle) and AB (60 ° rotation angle) stacking structures of bilayer graphene, the interface distance is 3.35 Å, which is close to those of the same bilayer graphene structures in references[38, 39]. As shown in Figure 2, the fat band structures and projected shells for four twisted bilayer graphene cells have been calculated along the path through *G-M-K-G*. The energy bands originated from the *s*, *p* and *d* orbitals are marked by red, blue and green lines, respectively. We obtained two Dirac cones around the *K*-point originated from *p* orbital because the band splitting occurs when two graphene layers interact. The splitting between the bands in the Dirac cones is responsible for low-dispersion bands near the *M* point[39, 40]. In our electron and phonon transport simulations, an overlapped TBGNRJ was built, which is divided into central part, left and right electrodes. The Brillouin zone of the junction is sampled by a 1×1×100 ***k***-mesh and a double zeta polarized for all atoms.

Table 1. Rotation angle ($\theta$), lattice constant ($a$), mean absolute strain ($\varepsilon$), and surface distance between two layers ($d$) for bilayer graphene.

| $\theta$ (°) | $a$ (Å) | $\varepsilon$ | $d$ (Å) |
|---|---|---|---|
| 0 | 2.46 | 0% | 3.35 |
| 21.8 | 6.51 | 0% | 3.25 |
| 38.2 | 6.51 | 0% | 3.25 |
| 60 | 2.46 | 0% | 3.35 |

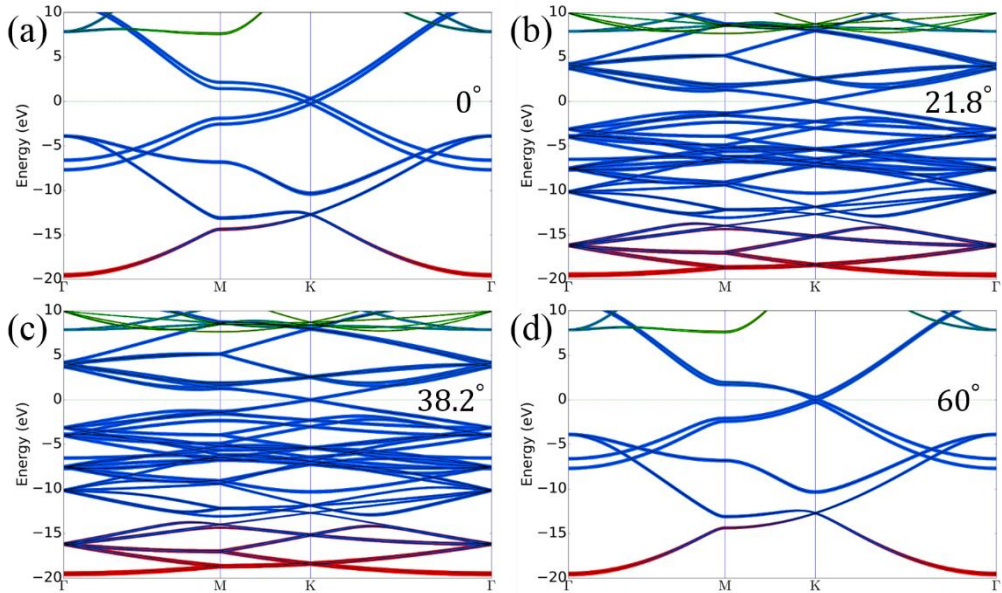

Figure 2. The computed fat band structures and projected shells of bilayer graphene structures with 0 ° (a) 21.8 ° (b), 38.2 ° (c) and 60 ° (d) rotation angles.

Based on these twisted bilayer graphene structures, overlapped TBGNRJs has been devised, followed by the investigation on electron transport and thermoelectric performance. Figure 3 shows the details of the device model, which is divided into central scattering region (SR), left electrode (LE) and right electrode (RE). Note the SR is composed of 5 unit cells along the transport direction because the *ZT* value is related to their junction length[21], while LE and RE are chosen from repetition matches.

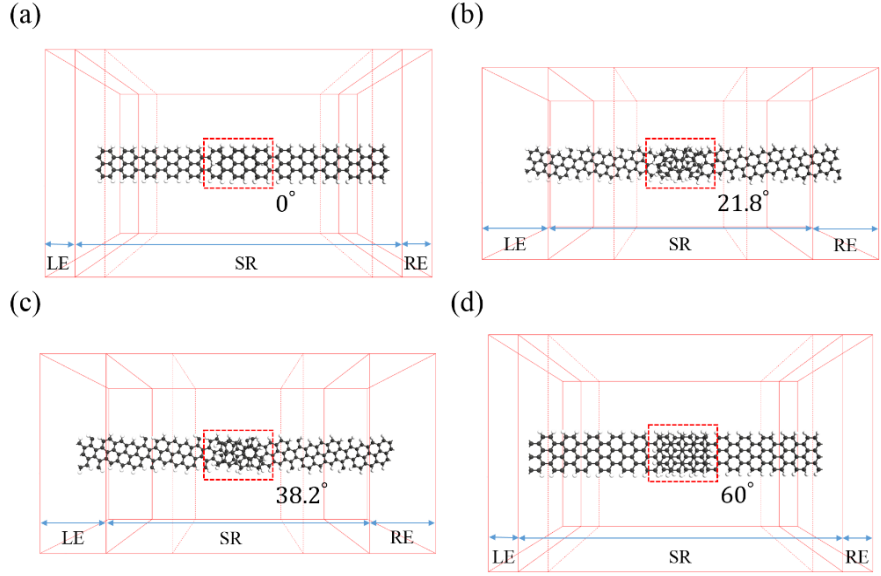

*Figure 3. The top view of the 0 ° (a) 21.8 ° (b), 38.2 ° (c) and 60 ° (d) TBGNRJs.*

## 3 Results and discussion

First, we investigate the *I-V* characteristics of these TBGNRs. The current across a TBGNR can be calculated from the Landauer-Buttiker equation[41]:

$$I = \frac{e}{h} \int dE \big(f_L(E) - f_R(E)\big) T_e(E) \tag{5}$$

Where, $T_e(E) = Tr[t^+ t](E)$ is the total electron transmission, $f_R(E)$ and $f_L(E)$ are the Fermi distribution functions of the right and left electrode, respectively.

Figure 4(a) shows the calculated current as a function of the bias voltage for the TBGNRJs in 0 °, 21.8 °, 38.2 ° and 60 ° rotation angles. With the increase of the bias voltage, the current displays an increasing trend. Moreover, the *I-V* curves of 0 °, 21.8 ° and 38.2 ° exhibit a behavior similar to a back-to-back p-n junction. However, it is not the case for the 60 ° TBGNRJ where the *I-V* curve exhibits transistor-like behavior. Similar results were shown in a prior publication[38]. At 21.8 ° and 38.2 ° rotation angles, the *I-V* and *dI/dV* curves (Figure 4(b)) are asymmetric, which is due to the non-central symmetric structure in the twisted junctions. Under the 1 *V* bias voltage, the current intensity of 21.8 ° TBGNRJ is higher than 38.2 ° TBGNRJ, while this relationship reverses at the -1 *V* bias voltage. Under the 0 ° and 60 ° rotation angles, the *I-V* and *dI/dV* curves exhibit symmetric behavior. For the *I-V* curve, under the same magnitude of the bias voltage, the current intensity of 0 ° TBGNRJ is smaller than 60 ° TBGNRJ for all voltages. For the *dI/dV* curves, the value of 60 ° TBGNRJ is higher than 0 ° TBGNRJ from -0.6 *V* to 0.6 *V*. However, 60 ° TBGNRJ is lower than 0 ° TBGNRJ at the rest voltages.

More interestingly, the *I-V* and *dI/dV* curves demonstrate NDR in 21.8 ° and 38.2 ° rotation angles under ±0.2 *V* bias voltage, which is mainly caused by transmission spectra under different bias voltages[42-44]. Figure 4(c) and (d) show the transmission spectra at 21.8 ° and 38.2 ° rotation angles under ±0.2 *V* and ±0.4 *V* bias voltage. The current under the bias voltage is mainly determined by the magnitude of transmission spectra. Compared with ±0.4 *V* bias voltage, the magnitudes of transmission spectra within ±0.2 *V* are significantly higher, resulting in a higher current intensity at ±0.2 *V* than ±0.4 *V*.

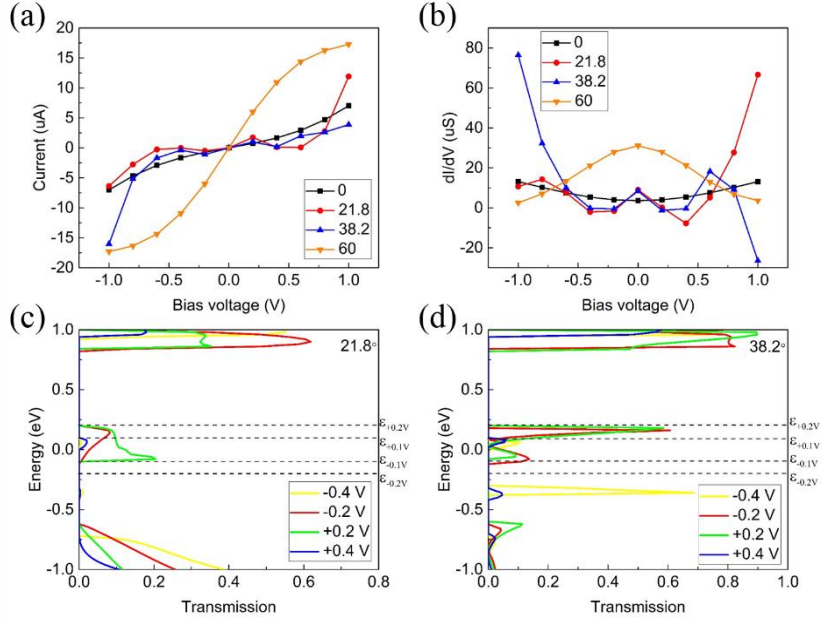

*Figure 4. The I-V (a) and dI/dV (b) curves of the TBGNRJs with 0 °, 21.8 °, 38.2 ° and 60 ° rotation angles. (c) and (d) are the transmission spectra of 21.8 ° and 38.2 ° TBGNRJs under ±0.2 V and ±0.4 V bias voltage.*

Thermoelectric properties of the TBGNRJs for the above twisting angles have also been simulated. Microscopically, the electron transport properties of the TBGNRJs can be described by the Boltzmann equation with the Fermi-Dirac distribution $f_{FD}(E,T)$. The electron conductance, Seebeck coefficient and heat transport coefficient of electrons are given by[45]:

$$G_e(T) = \frac{e^2}{h} L_0 \quad (6)$$

$$S(T) = \frac{1}{eT} \frac{L_1}{L_0} \quad (7)$$

$$\kappa_e(T) = \frac{L_0 L_2 - L_1^2}{ht L_0} \quad (8)$$

Where, *e* and *h* are the electron charge and Planck's constant.

$$L_n(T) = \int_{-\infty}^{+\infty} (E - E_F)^n T_e(E) \left(-\frac{\partial f_{FD}(E,T)}{\partial E}\right) dE \quad (9)$$

The heat transport coefficient of phonons is given by:

$$\kappa_{ph}(T) = \frac{1}{2\pi} \int_0^\infty \hbar w \, T_{ph}(\omega) \frac{\partial f_{BE}(\omega,T)}{\partial T} d\omega \quad (10)$$

Where, $f_{BE}(\omega,T)$ and $T_{ph}(\omega)$ are the Bose-Einstein distribution function and phonon transmission coefficient.

As shown in Figure 5, the electrical conductance, Seebeck coefficient, thermal conductance and *ZT* obtained in the TBGNRJs in the four twisting angles at 300 K are displayed. In the results, the chemical potential is a difference between the Fermi energy ($E_F$) and the DFT-predicted Fermi energy ($E_F^{DFT}$). The valleys and peaks of electrical conductance are shown in Figure 5(a) because of different band structures and bandgaps of the monolayer and bilayer parts of the TBGNRJs[21]. In Figure 5(b), the development of thermal conductance at four rotation angles have a similar trend with electrical conductance, which is similar to what is reported in the reference[16]. This result illustrates that the Seebeck coefficient is a key parameter to achieve high *ZT* value in the TGNRJs. In Figure 5(c), the maximum of Seebeck coefficient in 21.8 ° and 38.2 ° rotation angles is about 0.75 mV/K, which is about 7.5 times higher than the pristine graphene (~0.1 mV/K)[5, 46]. Figure 6 shows the band structures of the four TBGNRJs cells along the Brillouin path through *G-Z* because the Seebeck coefficient is derived from the bandgap[47]. Bandgap opening is shown following the twisting of the bilayer graphene, which results in an increase of the Seebeck coefficient[48]. As shown in Figures 6(b) and (c), the bandgap of the 21.8 ° and 38.2 ° TBGNRJs are 0.15 eV and 0.14 eV, which is larger than the bandgap of 0 ° (0.02 eV) and 60 ° (0.11 eV) TBGNRJs. Figure 5(d) shows the *ZT* values of four TBNGRJs, four peak *ZT* values are more than 1 for the 21.8 ° and 38.2 ° rotation angles at 300 K. Particularly, in the 21.8 ° rotation angle, the maximum *ZT* is 6.1, which is higher than most reported *ZT* values at room temperature[10, 49-51].

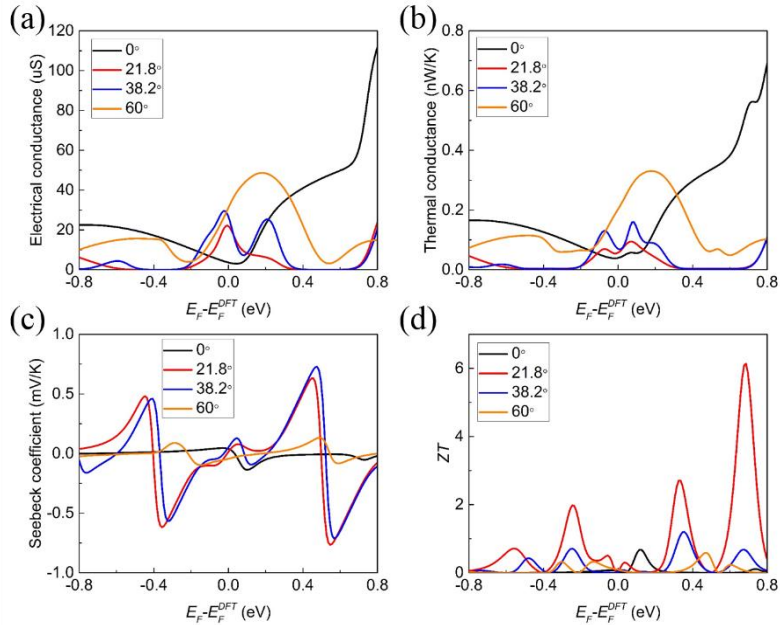

*Figure 5. The electrical conductance (a), thermal conductance (b), Seebeck coefficient (c) and ZT (d) obtained in the TBGNRJs with 0 °, 21.8 °, 38.2 ° and 60 ° rotation angles at 300 K.*

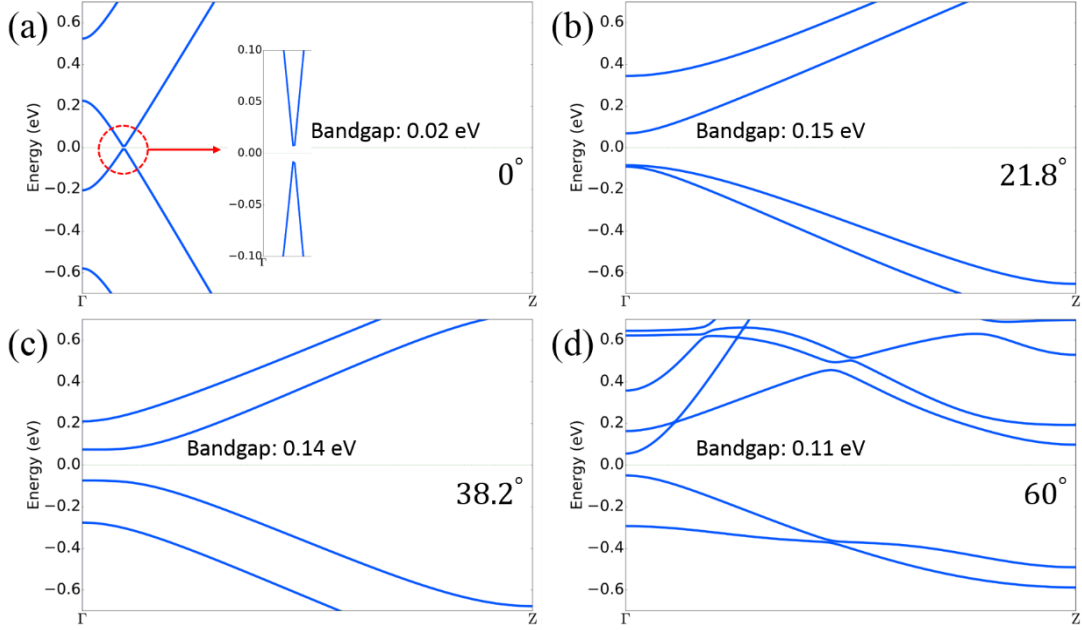

*Figure 6. The computed band structures of the TBGNRJs with 0 ° (a) 21.8 ° (b), 38.2 ° (c) and 60 ° (d) rotation angles.*

From Figure 5, the peaks of *ZT* higher than 1 exist in 21.8 ° and 38.2 ° rotation angles. We shall compute the thermal conductance of electrons and phonons for these TBGNRJs. As shown in Figure 7(a), the phonon contribution to the overall thermal conductance is much more than the electron contribution for *ZT* values of 2.7 and 2.0 in the 21.8 ° TBGNRJs. However, at the *ZT* value of 6.1, the electron contribution to thermal conductance is comparable to that from the phonon. The maximum *ZT* value of 38.2 ° TBGNRJs is 1.2, at which thermal conductance of electrons is slightly larger than that of phonons (Figure 7(c)). Figure 7(b) and (d) show that the thermal conductance of phonons increases with the rise of temperature. With all the electron and phonon transport properties calculated, we can evaluate the figure of merit of TBGNRJs. In Figure 5(d), all the outlines of *ZT* are rather asymmetric, and each of them has some peaks around the 0 eV. So, we can enhance *ZT* by appropriate *n*-type or *p*-type doping in these TBGNRJs at 300 K. For the 21.8 ° rotation angles, there are three peak values of *ZT* ~ 2.0, 2.7 and 6.1 at around -0.25 eV, 0.33 eV and 0.68 eV, which means *n*-type doping are more favorable than *p*-type doping to improve the thermoelectric performance in the 21.8 ° TBGNRJs. In the 38.2 ° TBGNRJs, the maximum *ZT* value is 1.2 at 0.33 eV, which can be obtained by *n*-type doping. All these results suggest that 21.8 ° TBGNRJs can obtain a significant improvement in thermoelectric performance at 300 K.

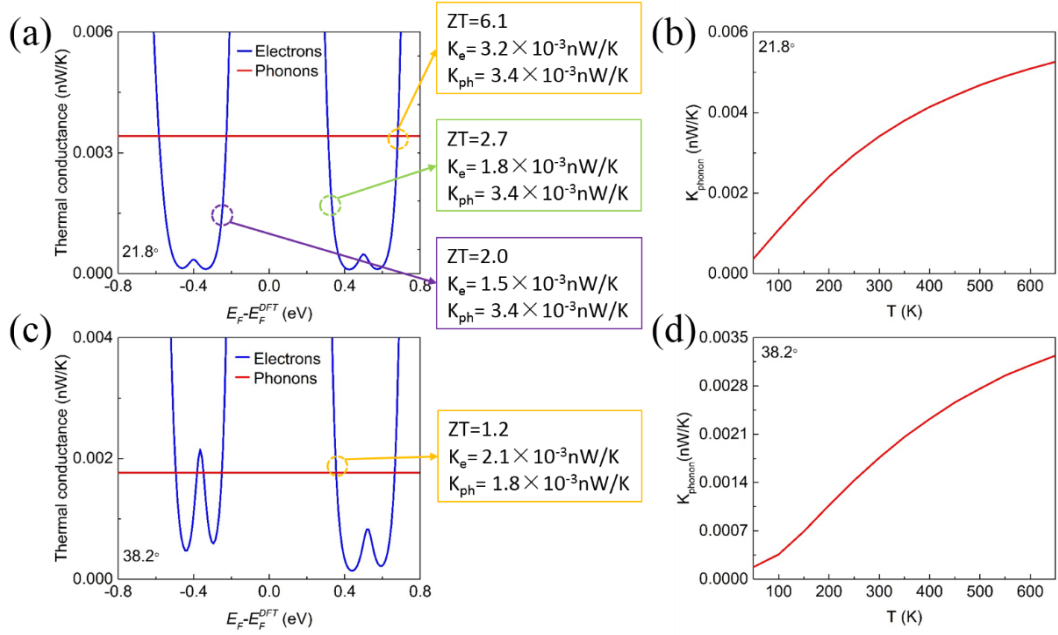

*Figure 7. The electron and phonon contribution to the thermal conductance in 21.8 ° (a) (b) and 38.2 ° (c) (d) TBGNRJs.*

## 4 Conclusion

We have investigated the electron transport and thermoelectric performance of TBGNRJs for 0 °, 21.8 °, 38.2 ° and 60 ° rotation angles by the first principles calculation. It is found that the *I-V* curves of 0 °, 21.8 ° and 38.2 ° rotation angles exhibit behavior similar to p-n junctions. However, the 60 ° TBGNRJs is different, which exhibits an *I-V* curve similar to a transistor. NDR is shown in TBGNRJs at 21.8 ° and 38.2 ° rotation angles under $\pm 0.2$ *V* bias voltage. High *ZT* values of 2.0, 2.7 and 6.1 have been achieved at -0.25 eV, 0.33 eV and 0.68 eV for the 21.8 ° rotation angles at 300K. It is interpreted that the reason of high *ZT* values of 21.8 ° TBGNRJ is due to its wider bandgap, which results in a much higher Seebeck coefficient. Moreover, the phonon contribution to thermal conductance is much more than that of electrons in *ZT* values of 2.0 and 2.7 in the 21.8 ° TBGNRJs. At the *ZT* value of 6.1, the electron contribution to thermal conductance is comparable to that of phonons. The outstanding *ZT* values of TBGNRJs make it a promising device structure for thermoelectric applications.

**Conflict of interest**
There are no conflicts to declare.

**Acknowledgements**
Authors are grateful for funding by the China Scholarship Council (CSC).

[41] M. Buttiker, Y. Imry, R. Landauer, S. Pinhas, Generalized Many-Channel Conductance Formula with Application to Small Rings, Phys. Rev. B 31(10) (1985) 6207-6215.
[42] K.M.M. Habib, S. Ahsan, R.K. Lake, Computational Study of Negative Differential Resistance in Graphene Bilayer Nanostructures, Proc Spie 8101 (2011).
[43] J. Kumar, H.B. Nemade, P.K. Giri, Density functional theory investigation of negative differential resistance and efficient spin filtering in niobium-doped armchair graphene nanoribbons, Phys. Chem. Chem. Phys. 19(43) (2017) 29685-29692.
[44] D. Zhang, M.Q. Long, X.J. Zhang, L.L. Cui, X.M. Li, H. Xu, Perfect spin filtering, rectifying and negative differential resistance effects in armchair graphene nanoribbons, J. Appl. Phys. 121(9) (2017).
[45] H. Sadeghi, S. Sangtarash, C.J. Lambert, Oligoyne Molecular Junctions for Efficient Room Temperature Thermoelectric Power Generation, Nano Lett. 15(11) (2015) 7467-7472.
[46] J.G. Checkelsky, N.P. Ong, Thermopower and Nernst effect in graphene in a magnetic field, Phys. Rev. B 80(8) (2009).
[47] Y. Yokomizo, J. Nakamura, Giant Seebeck coefficient of the graphene/h-BN superlattices, Appl. Phys. Lett. 103(11) (2013).
[48] J.M. Zheng, P. Guo, Z.Y. Ren, Z.Y. Jiang, J.T. Bai, Z.Y. Zhang, Conductance fluctuations as a function of sliding motion in bilayer graphene nanoribbon junction: A first-principles investigation, Appl. Phys. Lett. 101(8) (2012).
[49] H.L. Liu, X. Shi, F.F. Xu, L.L. Zhang, W.Q. Zhang, L.D. Chen, Q. Li, C. Uher, T. Day, G.J. Snyder, Copper ion liquid-like thermoelectrics, Nat. Mater. 11(5) (2012) 422-425.
[50] L. Han, D.V. Christensen, A. Bhowmik, S.B. Simonsen, L.T. Hung, E. Abdellahi, Y.Z. Chen, N.V. Nong, S. Linderoth, N. Pryds, Scandium-doped zinc cadmium oxide as a new stable n-type oxide thermoelectric material, J Mater Chem A 4(31) (2016) 12221-12231.
[51] R. Chetty, A. Bali, R.C. Mallik, Tetrahedrites as thermoelectric materials: an overview, J. Mater. Chem. C 3(48) (2015) 12364-12378.

# Enhanced thermoelectric performance of twisted bilayer graphene nanoribbons junction

Shuo Deng[1,2], Xiang Cai[2], Yan Zhang[3*], and Lijie Li[2*]

[1]School of Logistic Engineering, Wuhan University of Technology
[2]College of Engineering, Swansea University, Swansea, SA1 8EN, UK
[3]School of Physics, University of Electronic Science and Technology of China
*Emails: zhangyan@uest.edu.cn, L.Li@swansea.ac.uk

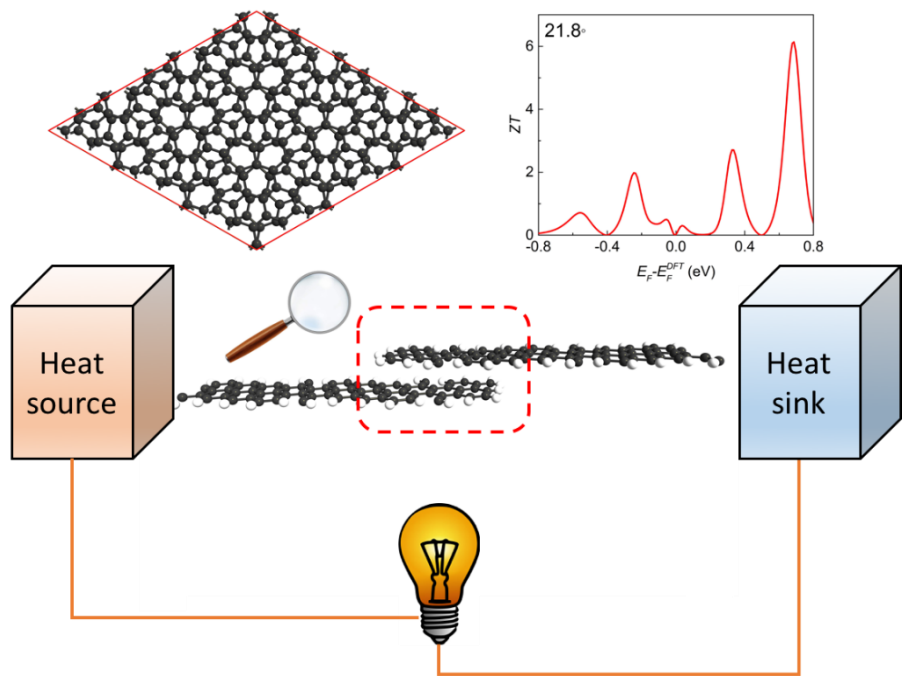